# *Ex vivo* experiment on vertebral body with defect representing bone metastasis

W. Lokbani[a,b], V. Allard[a,b], T. Broussolle[c], CY. Barrey[c,d], C.B. Confavreux[b,d], K. Bruyère[a], JP. Roux[b], F. Bermond [a], H. Follet[b], D. Mitton[a]

[a] *Univ Lyon, Univ Gustave Eiffel, Univ Claude Bernard Lyon 1, LBMC UMR_T9406, 69622 Lyon, France*

[b] *Univ Lyon, Univ Claude Bernard Lyon 1, INSERM, LYOS UMR 1033, 69008 Lyon, France*

[c] *Service de Chirurgie du Rachis et de la Moelle Epinière, Hôpital Neurologique et Neurochirurgical, GHE, Hospices Civils de Lyon, Université Claude Bernard Lyon 1, Lyon, France*

[d] *Centre Expert des Métastases Osseuses (CEMOS) , Département de Rhumatologie, Hôpital Lyon Sud, Hospices Civils de Lyon, Lyon, France;*

**Corresponding authors:**

David Mitton, david.mitton@univ-eiffel.fr, ORCID number: 0000-0001-9260-0916

Helene Follet, helene.follet@inserm.fr, ORCID number: 0000-0002-3290-2899

**Running title:** Dataset for human vertebral analysis

**Word count**: Abstract: 214, Main text: 3665

## Abstract

*Background* Osteolytic metastases located in the vertebrae reduce strength and enhance the risk of vertebral fractures. This risk can be predicted by means of validated finite element models, but their reproducibility needs to be assessed. For that purpose, experimental data are requested. The aim of this study was to conduct open-access experiments on vertebrae, with artificial defect representing lytic metastasis and using well-defined boundary conditions. *Methods* Twelve lumbar vertebral bodies (L1) were prepared by removing the cortical endplates and creating defects that represent lytic metastases, by drilling the cancellous bone. Vertebral bodies were scanned using clinical High Resolution peripherical Quantitative Computed Tomography before and after defect creation for 3D reconstruction. The specimens were then tested under compression loading until failure. Surface Digital Image Correlation was used to assess strain fields on the anterior wall of the vertebral body. *Findings* Mean failure load was 3.30 kN *(range 1.46-7.67 kN)* and stiffness was 13.64 kN/mm *(range 3.36 – 40.94 kN/mm).* At failure, mean values of equivalent von Mises strains and minimum principal strains were 1.53% *(range 0.51 – 2.55 %)* and -1.43% *(range -2.34 – -0.35 %)*, respectively. *Interpretation* These data (biomechanics data and the tomographic images needed to build subject-specific models) are shared with the scientific community in order to assess different vertebral models on the same dataset.

## 1. Introduction

More than 18 million cancers are registered each year (Coleman et al., 2020b). Tumor cells can migrate from their primary tumor site to another location leading to a secondary cancer called metastasis. Bone is the third most frequent site of metastases behind lung and liver (Macedo et al., 2017). Bone metastases are highly prevalent in patients with breast, prostate and lung cancer with an incidence of 47% to 85%, 33% to 85% and 30% to 55%, respectively (Coleman et al., 2020a). Since vertebrae are highly vascularized, they are often affected, especially in the posterior part of the vertebral body (Algra et al., 1992). Bone metastases located in vertebrae can induce severe pain, immobilization and reduce quality of life. From a mechanical point of view the main issues of bone metastases are the risk of pathologic fracture (10-30% of all cancer patients) and spinal cord compression (5%) (Selvaggi and Scagliotti, 2005) which can lead to paraplegia. Currently, clinicians use the Spinal Instability Neoplastic Score (SINS) to assess stability and the risk of vertebral fracture (Fisher, 2010). This score ranges from 1 to 18 based on several criteria: the location of the lesion, its osteolytic nature, the existence of mechanical pain, radiographic alignment, vertebral body collapse and posterolateral involvement. Although this tool correctly predicts extreme cases, it remains unsatisfactory for intermediate values (score between 7-12) (Shi, 2018).

To overcome SINS limitations, finite element models (FEMs) have been developed to evaluate the strength of vertebrae and predict the risk of mechanical failure (Chevalier, 2008a; Choisne et al., 2018; Costa et al., 2020). But only a few studies have presented experimentally validated finite element models for strength assessment of vertebrae with defects, which simulate vertebral metastases (Anitha et al., 2017; Stadelmann, 2020).

The experimental data for model validation are limited in number and few metastatic vertebrae are available. Thus, lytic metastatic defects were simulated by creating defect in vertebrae. Various techniques have been used to create such defects. The main options are through the cortical shell of the vertebral wall (McGowan et al., 1993), through the pedicles (Palanca et al., 2018) or vertebroplasty balloon (Giambini et al., 2016).

Drilling through the vertebral wall causes cortical damage (Alkalay et al., 2018; Alkalay, 2015; Groenen, 2018; Silva et al., 1993; Whealan et al., 2000; Whyne et al., 2003) which leads to vertebral strength reduction (Silva et al., 1993).

Among protocols leading to vertebral fracture, Dall'Ara et al. developed an experimental protocol for inducing an anterior wedge-shape fracture (typically osteoporosis fractures) in intact vertebral bodies (Dall'Ara, 2010). In addition, a recent study used the same protocol for validating a numerical model on real metastatic vertebrae (Stadelmann, 2020). An interesting point of this protocol was removal of the endplates and a loading test done directly through the cancellous bone to obtain well-defined boundary conditions. Thus, no embedding material was inserted between the vertebra and the machine. Removal of the vertebral endplates gives a direct access to the cancellous bone to create a defect without cortical damage of the vertebral wall. This approach allows the creation of a defect representing a lytic metastasis and the definition of well controlled boundary conditions to further assess finite element models.

Most of the time, the experimental data could not be simply used by other teams to build subject-specific models, because the images are not in open access. Therefore, there is a need for open data (from images up to biomechanical tests) for vertebral bodies representing metastatic lesions.

In this context, the aim of this study was to perform experiments on vertebral bodies with defect simulating a worth lytic metastasis and with controlled boundary conditions. These experimental data (images and biomechanical experiments results) are available as open data (*https://doi.org/10.25578/KSCRGA).*

## 2. Methods

### *2.1. Sample selection*

Twelve fresh frozen lumbar vertebrae (L1) were obtained from the Anatomy Department of Rockefeller University (DUAR Lyon, France, authorization no. DC-2015-2357). The test group was composed of 5 female and 7 male donors aged 64 – 95 years (mean = 83). Soft tissues and intervertebral discs were removed, vertebrae were kept hydrated in saline-soaked gauze and stored at –20 °C in vacuum-sealed bags.

Vertebrae were imaged using a clinical Quantitative Computed Tomography (QCT) scan (GE Medical Systems) equipped with a calibration phantom of five different densities (QCTPro CT Calibration Phantom, Mindways, Austin, USA). Each vertebra was scanned according to an *in vivo* protocol (Voltage: 120 kV, Current intensity: 270 mA, matrix size 512 x 512, Field Of View (FOV) reconstruction: FOV 20cm (Voxel size: 0.390 x 0.390 x 0.625 $mm^3$) and FOV 36cm (Voxel size: 0.703 x 0.703 x 0.625 $mm^3$), Standard “B” filter). Density was computed from calibration according to the equation (1) :

$$\text{Density (in g } K_2HP0_4/cm^3) = 0.0006855 \text{ greylevels} + 0.01745 \qquad (1)$$

Before defect creation and for comparison purpose, bone mineral density (BMD, in $g/cm^3$) was also obtained using HRpQCT (High Resolution peripherical Quantitative Computed Tomography, effective energy of 60 kVp, x-ray tube current of 95 mA, and matrix size of 1536 x 1536, Voxel size: 0.082 x 0.082 x 0.082 mm). Attenuation data were converted to equivalent hydroxyapatite (HA) densities. Quality control, based on Shewart rules, was monitored by daily scans of a phantom containing rods of HA (densities of 0, 100, 200, 400, and 800 mg $HA/cm^3$) embedded in a soft-tissue equivalent resin (QRM, Moehrendorf, Germany) (Boutroy et al., 2005). This calibration gave the equation (2) for the density computation :

$$\text{Density (in g HA/ cm3)} = 0.0001778 \text{ greylevels} - 0.358 \qquad (2)$$

*Sample preparation*

First, an identification of endplates was done for each vertebra based on QCT images. A distance was defined between the upper or lower extremity of the vertebra and a plane in the cancellous bone respectively below and above the vertebral endplates.

Then, posterior parts of the vertebrae were embedded in a cubic mold filled with resin to retain the vertebra. To cut the endplates, the diamond blade of the saw (Isomet 4000, Buehler) was placed in contact with the vertebral endplate and moved to the distance defined on the QCT images. This approach helped to cut the endplates in a personalized way. Cuts were performed under constant water irrigation **(Fig. 1.)**.

### *2.2. Defect creation*

After cutting the endplates, defects were created while the vertebra was retained by its embedded posterior part. Defects were realized using a column drill (PROMAC 210Z) closer to the posterior part of the vertebral body (Fig. 2.) which is the main location for vertebral metastases (Algra et al., 1992). Based on Taneichi et al. (1997), defects represented around 30% of the middle transversal cross-section of the vertebral body to ensure the case of impending vertebral collapse (instability). To reach this target, defect size and shape were defined from QCT images and expected defect contours were highlighted on the vertebral body as a template for drilling an adequate defect without causing cortical damage of the vertebral wall. Defects were drilled from the superior surface of the vertebral body until its inferior surface, using a mounted diamond core drill (∅8 mm, Starlite, Bryn Mawr, PA 19010, USA). This final shape of the defect (Fig. 2D) was chosen for its reproducibility and as a simplification of the metastasis's shapes in the vertebral body. The embedded posterior part of the vertebra was removed by cutting the pedicles and keeping only the vertebral body without endplates for the mechanical tests. As for the cutting of the endplates, the vertebral body was rinsed by constant irrigation during the cutting process of the pedicles. Soft tissues were removed from the anterior vertebral wall to apply the random speckle needed for the strain field measurement using Digital Image Correlation (DIC). Twelve measurements of vertebral bodies heights (L) were performed around the perimeter to assess the parallelism between the two surfaces (Dall'Ara, 2010).

Finally, post-defect QCT scans were performed on the vertebral bodies to compute defect size ratio based on 2D and 3D measurements (**Fig. 3.)**. The specimens were scanned with the same parameters mentioned previously.

The post-defect QCT data and the pre-defect HRpQCT data were combined to build the 3D model of the damaged vertebrae in view of finite element modelling. QCT density values do not represent true density value even with the use of a calibration phantom, a possible hypothesis is the presence of air in the cancellous bone that may have influenced the density acquisition (Fig. 3). Thus, we decided to use the segmentation of the post-defect QCT images to get the geometry combined with the pre-defect HRpQCT images to recover density values. To locate the post-defect segmentation into the pre-defect scan, the cortical shells from segmentations from both scans were fitted using a coherent point drift algorithm (PyCPD, *https://pypi.org/project/pycpd*). The transformation matrix between segmentations from QCT and HRpQCT scans was calculated. This resulted in a post-defect QCT mask applied to the pre-defect HRpQCT scan. In order to mimick the QCT scan quality, the HRpQCT scan was also resampled to the QCT scan resolution (0.410 x 0.410 x 0.410 $mm^3$). So, we were able to access to HRpQCT grey levels corresponding to the post defect QCT segmented volume.

### *2.3. Mechanical tests*

All specimens were first thawed at 4°C over 12 h, then kept wrapped in wet gaze at room temperature for 2 h before the mechanical tests (Alkalay et al., 2018). The center of mass of each intact vertebral body was computed from QCT images and projected in the inferior slice (Dall'Ara, 2010; Stadelmann, 2020). Vertebral body contours of this slice, the projected center of mass and the frontal and sagittal axis of the vertebral body were printed and placed on the testing system in order to align the center of mass with the compression axis of the machine.

Vertebral bodies were compressed between two steel plates covered by sandpaper **(Fig. 4.)** to avoid specimen sliding during the compression test (Dall'Ara, 2010; Stadelmann, 2020). Uniaxial compression with a ball joint (lower platen) was chosen to induce a crush fracture which is the major type of metastatic fracture (McCloskey, 1993).

Steel plates were mounted in an electromagnetic testing machine (Instron ElectroPuls E10 000, England). First, a preload of 25 N was applied to ensure the contact between the superior plate and the specimen (Stadelmann, 2020). Then, uniaxial compression was applied at a rate of 5 mm/min until failure (Chevalier, 2008b; Dall'Ara, 2010). The load was measured using a 10 kN load cell (Instron, Serial No.: 107 182) with an accuracy of 1 N. The compression of vertebral bodies was computed from the testing machine displacement (internal sensor Heidenhain, Serial No. 201003, accuracy of 0.01 mm), corrected by the compliance of the mechanical set-up that was measured previously. The outcomes of each test were the load-displacement curve and the strain field measured on an area of interest located on the anterior wall of the vertebral body. We defined failure loads as the peak force on the load-displacement curve, and the compression at failure as the corresponding displacement. Stiffness (k, in kN/mm) was computed as the slope of the linear part of the load-displacement curve between 25% and 75% of failure load (Wegrzyn, 2011). Apparent Maximal Stress was calculated as the ratio of Maximal load on apparent cross-section (F/Sapp), and the Apparent Modulus was calculated as the ratio of stiffness*vertebrae's height over apparent cross-section (k*L/Sapp).

### *2.4. Strain field measurements*

Strain fields were measured on the anterior wall of the vertebral body by performing stereo Digital Image Correlation (DIC) using VIC-3D software (VIC3D 2010, Correlated Solutions Inc.). The anterior wall of the vertebral body was covered with a thin layer of white clown makeup (Aquacolor, Germany) and black paint was sprayed on the white background to provide randomized speckles (Revel et al., 2022). Two Photron FASTCAM SA3 video cameras with 105 mm lenses Macro (EX Sigma for Nikon A FD) recorded the test at a frame rate of 50 images per second. The distance between the two video cameras was 150 mm and the stereo angle was 10°. The distance to specimen was 420 mm. Two projectors (Dedolight 400 D) provided sufficient lighting to avoid any shadow areas on the images and to ensure a good contrast.

The system was calibrated with a grid (12x9 points, spaced 5mm). The mm/pixel ratio was equal to 0.07. For DIC, the subset size was chosen equal to 35 px (physical size of 2.45mm), the step equal to 15 px (physical size of 1mm) and the filter size was 15. The Area Of Interest (AOI) located on the anterior wall of the vertebral body was around 20 x 10 $mm^2$, depending on the vertebral body size. All the specimens were first captured in an unloaded state to get the zero-strain configuration and to assess the strain uncertainties (Palanca et al., 2018). To assess the DIC quality, the sigma error can also be considered: for all pairs of images, it was lower than 0.02px in the major area of the AOI.

To illustrate the Green-Lagrange strain fields, the maximum and minimum principal strains ($\varepsilon_1$, $\varepsilon_2$), and the equivalent von Mises strains ($\varepsilon_v$) (Equation 3) are given.

$$\varepsilon_v = \frac{2}{3}\sqrt{\varepsilon_1^2 - \varepsilon_1\varepsilon_2 + \varepsilon_2^2} \qquad (3)$$

### *2.5. Statistics*

A Shapiro-Wilk test was performed to assess the normality of the data. Scatterplots are used to present the comparison with literature between failure load and stiffness. Pearson regression coefficients were used. The significance level p was set to 0.05. A Bland and Altman analysis was done to compare the density measurements from QCT and HRpQCT images.

## 3. Results

Descriptive characteristics are presented in **Table 1**. Good agreement was found between the two modalities QCT and HRpQCT for BMD measurement (mean difference from Bland and Altman analysis equal to 0.05 $g/cm^3$).

Load-displacement curves obtained for all specimens are presented in **Fig. 5.** The values of failure load, displacement at failure load, stiffness, mean principal strains and mean equivalent von Mises strains at failure are presented in **Table 2**. The mean value of failure loads was 3.30 kN ± 1.62 (range 1.46 – 7.67 kN) with a mean displacement of 0.7 mm ± 0.2 (range 0.5 – 1.1 mm). Mean stiffness was 13.64 kN/mm ± 10.23 (range 3.36 – 40.94 kN/mm). A linear correlation was found between failure load and stiffness **(Fig. 6.)** with a determination coefficient $r^2$ = 0.98. Apparent Max. Stress was 4.35 MPa ± 1.87 (range 1.91 – 8.32) and Apparent Modulus was 249 MPa ± 145 (range 75 – 586).

Strain field measurements qualitatively highlighted the compression of the anterior wall even if some lack of symmetry can be observed **(Fig. 7.)**. Considering the zero-strain configuration, systematic errors and random errors were 0.009% and 0.001%, respectively. On the last configuration recorded before failure, the mean values of equivalent von Mises strains and minimum principal strains on the AOI were in average equal to 1.53% (range 0.51 – 2.55 %) and -1.43% (range -2.34 – -0.35%), respectively.

For the defect size ratio, there was no significant difference (p = 0.858) between the estimation from the 2D middle transversal cross-section of the vertebral body and from the 3D measurement on the whole vertebral body.

## 4. Discussion

The aim of this experimental study was to quantify failure loads and strain fields of vertebral bodies with defects under uniaxial compression in order to serve as an open database. The current data until failure are complementary to open data on human spine segments loaded under compression-bending (http://amsacta.unibo.it/6977/).

To investigate the influence of defects on vertebral strength, and to be able to compare the experimental results to numerical metrics, we decided to reproduce the experimental protocol without endplates as done in (Dall'Ara, 2010) or (Stadelmann, 2020). Nevertheless, some points of the experimental protocol needed to be adapted. The modifications were done first regarding the polishing. In previous studies, after cutting the endplates with a diamond band saw, vertebral surfaces were not perfectly parallel (Dall'Ara, 2010; Stadelmann, 2020) and they polished vertebral surfaces. In the current study, we decided to cut the endplates with a diamond blade to obtain a clean-cut surface. However, as the vertebrae were weakened by the defect creation, we chose not to polish vertebral surfaces, even if this step could be done smoothly. The second modification was related to the hydration before the mechanical test. In previous studies (Dall'Ara, 2010; Stadelmann, 2020), each specimen was kept in a 0.9% saline solution for 1 h before the mechanical tests. In the current study, according to the conservation of our specimens (hydrated in saline-soaked gauze and stored at –20 °C in vacuumed plastic bags), we kept the vertebrae in saline-soaked gauze before the mechanical tests. Moreover, endplate cutting was done under constant irrigation. In addition, we considered that the presence of liquid around the vertebrae would have compromised the digital image correlation for strain field measurements during the compression test. The third modification was related to the positioning of the vertebrae. Both previous studies (Dall'Ara, 2010; Stadelmann, 2020) shifted the center of mass of the vertebrae from the loading axis in order to have an anterior wedge fracture (specific to osteoporotic fractures). In our case, we focused on metastatic fractures on lumbar vertebrae (L1) mainly defined by a crush deformity (McCloskey, 1993). Consequently, the hypothesis of uniaxial compression was consistent.

The results of the current study, on vertebrae with artificial lesions, are compared with those of (Dall'Ara, 2010) and (Stadelmann, 2020) obtained, respectively, on intact vertebrae and real metastatic vertebrae (**Table 3)**.

The failure loads measured here were lower than those found in the literature, which is consistent due to the removal of cancellous bone to create the defect. The range of failure loads in (Stadelmann, 2020) are much higher than other studies because three in four L1 specimens had osteoblastic metastases, thus the failure load was significantly higher than specimens with osteolytic metastases.

From Dall'Ara et al, mean stiffness, calculated as the slope of the linear part of the ''load–axial displacement'' curve, in intact vertebrae was 33.52 kN/mm (unpublished data) compared to 13.64 kN/mm in our study. This huge difference can be explained by the state of the vertebral bodies (intact *vs.* damaged). In addition, a variation could be related to the stiffness computation. In the current study, stiffness was computed by the slope in the linear part of the load-displacement curve in the interval corresponding to 25% - 75% of failure load (Wegrzyn, 2011).

In the current study, failure load and stiffness shown a linear correlation ($r^2$ = 0.98) for 12 tested specimens. For intact vertebrae, a linear correlation was also found ($r^2$ = 0.69) on 8 tested specimens by Dall'Ara et al. **(Fig. 6.)**.

Strains were already assessed by DIC on intact vertebral bodies (Gustafson et al., 2017). They found that, prior to reaching the global yield force, more than 10% of the DIC measurements reached a minimum principal strain of -1.0%. In the current study, the minimum principal strain was –1.43% at the last configuration recorded before failure. Strains were also measured on vertebral segments with simulated defects (Palanca et al., 2018) and with real metastases (Palanca, 2021) under different loading conditions. In the current study, we studied only vertebral bodies without endplates, posterior parts and vertebral discs. Moreover, we tested the specimens in compression until failure and we focused on strains at failure while Palanca et al. investigated the strains for different loads corresponding to a given compression level, on intact vertebra. They analyzed how the metastatic vertebra deformed compared to the intact vertebrae (Palanca, 2021).

To choose a defect size ratio, the value of 30% based on the work by Taneichi et al. (1997) was considered. The defects represented around 30% of the 2D middle transversal cross-section of the vertebral body to ensure the case an impending vertebral collapse (instability) based on QCT images. Numerous studies investigated defect size influence on vertebral strength. Nevertheless, no consensus was found about the basic definition of defect size. Some studies considered defect size as the cross-sectional area of a defect divided by the cross-sectional area of the vertebral body (Alkalay et al., 2018; McGowan et al., 1993; Silva et al., 1993; Taneichi et al., 1997; Whealan et al., 2000). The cross-sectional area was either in the midplane of the vertebral body (McGowan et al., 1993; Silva et al., 1993), in the largest lytic cross-sectional area within the vertebral body (Alkalay et al., 2018), or the average of several cross-sections (Taneichi et al., 1997). Others considered the whole vertebral volume (Giambini et al., 2016; Whyne et al., 2003; Windhagen et al., 1997). In the current study, after creation of the defect, vertebral bodies were rescanned, allowing the determination of the exact 3D defect volume in addition to the assessment of the defect size ratio based on the 2D middle transversal cross-section. As expected for an end-to-end defect, no difference was found between the 2D and 3D measurements; thus, the assessment of the defect size ratio, in this specific case, can be made from the middle transversal cross-section.

Finally, this study has some limitations. First, these experiments are different from clinical cases. They were performed on isolated vertebral bodies, without their endplates, posterior arches, disco-ligamentary structures or adjacent vertebrae. Moreover, the measured strain fields are not volumetric but surfacic, and considering the variability of vertebral body external shape and the video configuration (one pair of video cameras), strain fields were assessed partially on the cortical wall. However, this study offers new data, obtained in well-defined boundary and loading conditions, that are complementary to existing open data (http://amsacta.unibo.it/6977/).

## 5. Conclusion

In the current study, we conducted experiments in controlled conditions based on previous studies, to assess vertebral strength and strains at failure, in the specific case of vertebral bodies with defects mimicking lytic metastases. Defects mimicking lytic metastases were created in vertebral body without affecting the cortical of vertebral walls. Well controlled boundary conditions were chosen, and surface strains were computed by digital image correlation to assess local behavior. The data obtained will be useful for the building of subject-specific models and for their validation, which is a step needed to go towards clinical applications of numerical prediction of bone failure. These data are given in open access to further improve the credibility of the models (https://doi.org/10.25578/KSCRGA).

## Acknowledgement

The authors would like to acknowledge Leila Ben Boubaker for her assistance during the experiments, Stéphane Ardizzone, and Richard Roussillon for their technical support. Cynthia Goutain-marjorel and Catherine Planckaert for their dedication and help during the image acquisition.

Pr. Philippe Zysset for discussion on the adaptation of the current protocol regarding the previous studies. Dr Enrico Dall'Ara and Dr Marco Palanca for sharing unpublished data.

This work was partly funded by LabEx Primes (ANR-11-LABX-0063) and MSDAVENIR Research Grant (CBC).

## Contributor's statement

**Wyssem Lokbani**: Methodology, Data curation, Writing -original draft preparation **Valentin Allard**: Data curation, Writing- Reviewing **Théo Broussolle**: Methodology, Writing- Reviewing **Cédric Barrey**: Conceptualization, Supervision, Writing- Reviewing **Cyrille Confavreux**: Conceptualization, Supervision, Writing- Reviewing, Funding acquisition **Karine Bruyère**: Methodology, Data curation, Writing- Reviewing **Jean-Paul Roux**: Methodology, Writing- Reviewing **François Bermond**: Methodology, Data curation, Writing- Reviewing **Hélène Follet**: Conceptualization, Supervision, Data curation, Writing- Reviewing, Funding acquisition **David Mitton**: Conceptualization, Methodology, Supervision, Writing- Reviewing, Funding acquisition

## Conflict of interest

None

## Figure legends

**Fig. 1.** Sample preparation steps: A) Fixation of the posterior part of the vertebra in the resin. B) Cutting endplates with a diamond blade. C) Lateral view of cutting endplates step. D) Superior surface of the vertebral body after cutting the endplate. E) Inferior surface of the vertebral body after cutting the endplate

**Fig. 2.** Defect creation steps: A) Positioning the posterior vertebral part in the vice before drilling. B) Defect created by drilling. C) Final vertebral body free of endplates, with defect and cleaned of the residual soft tissue. D) top view of the vertebral body after smoothened the defect edges with a scalpel

**Fig. 3.** Quantitative Computed Tomography images from left to right, in the transversal, frontal and sagittal planes. (A) Intact vertebra and (B) vertebral body with defect

**Fig. 4.** Mechanical test set up: A) Positioning the template of the tested vertebral body on the sandblasted surface. Vertebral body axes (frontal and sagittal) were aligned with machine axes. B) Vertebral body with contrasted speckle ready for compression and video acquisitions. The lower platen is mounted via a ball joint C) General view of the set up with the two video cameras.

**Fig. 5.** Load-Displacement curves of all the specimens (each specimen number corresponding to the curve is mentioned by " # specimen number ")

**Fig. 6.** Failure load-Stiffness correlation in the current study *(vertebral bodies without endplates and artificial defects)* and in (Dall'Ara, 2010) *(intact vertebral bodies without endplates)*

**Fig. 7.** Example of (left) Mean maximum Green-Lagrange principal strain $\varepsilon 1$, Mean minimum Green-Lagrange principal strain $\varepsilon 2$ (microstrain) at failure and, (Right) Equivalent von Mises strain (microstrain) fields on the anterior wall of the vertebral body at failure load *(specimen #9)*

## Table legend

**Table 1.** Initial characteristics of the specimens: Sample ID, Donor sex, Donor age, Bone Mineral Density (BMD) defined by HRpQCT before creating the defects, BMD defined by QCT before and after creating the defects, Bone volume ratio (BV/TV) before creating defect, Mean height of the vertebral body after cutting the endplates (11 measurements taken around each vertebral body) , Defect size ratio measured in 2D (from one QCT cross-section taken at the middle of the vertebral body), Defect size ratio measured in 3D (from QCT).

**Table 2.** Overview of the results from the mechanical tests: Sample ID, Failure load, Compression at failure load, Stiffness, Mean equivalent von Mises strain at failure, Mean maximum Green-Lagrange principal strain $\varepsilon 1$, Mean minimum Green-Lagrange principal strain $\varepsilon 2$ (compressive strains) at failure, Apparent Stress and Apparent Young's Modulus.

**Table 3.** Comparison of failure loads and stiffnesses with the literature. Means, medians, and ranges are indicated.

**Fig. 1.**

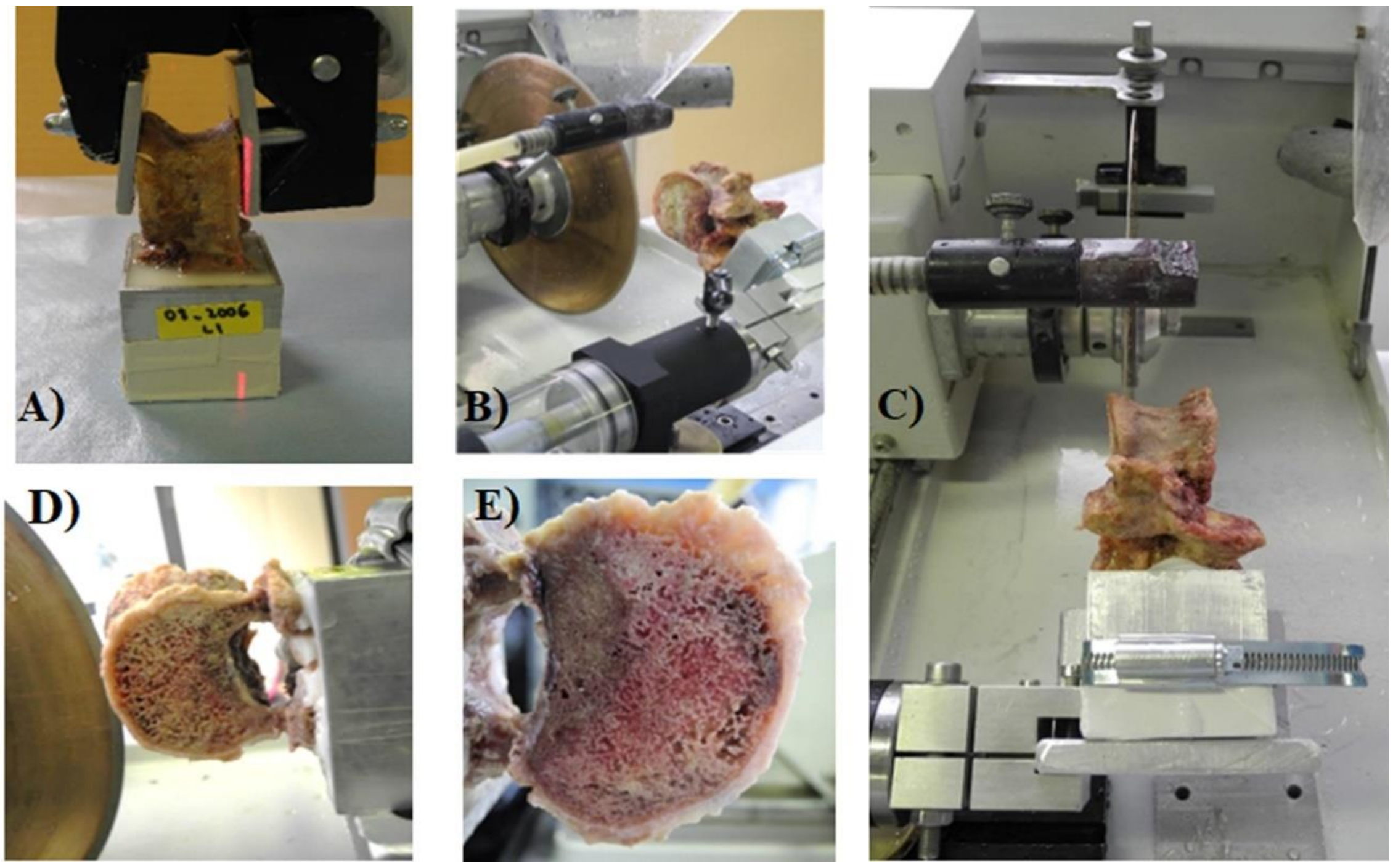

**Fig. 2.**

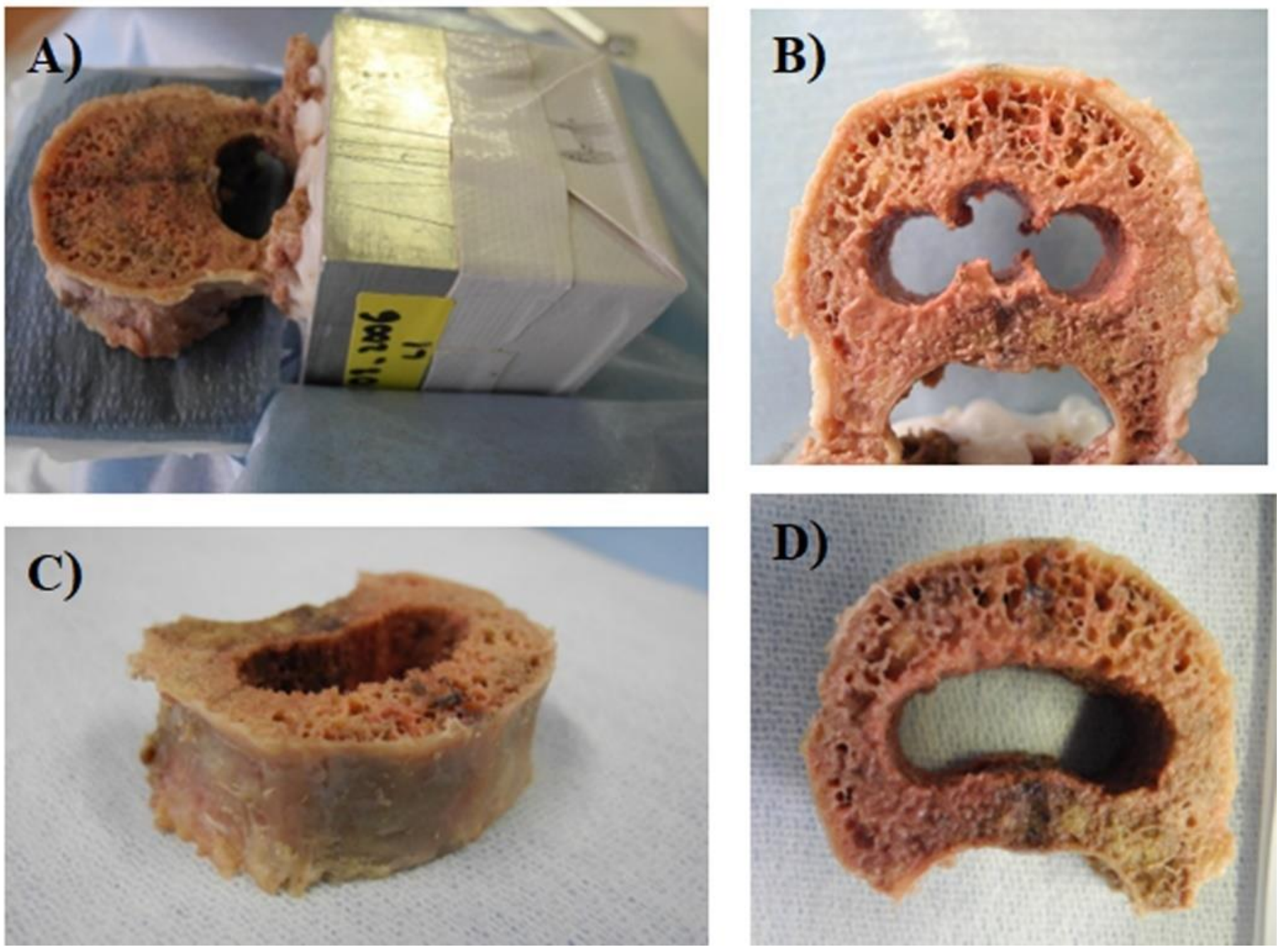

**Fig. 3.**

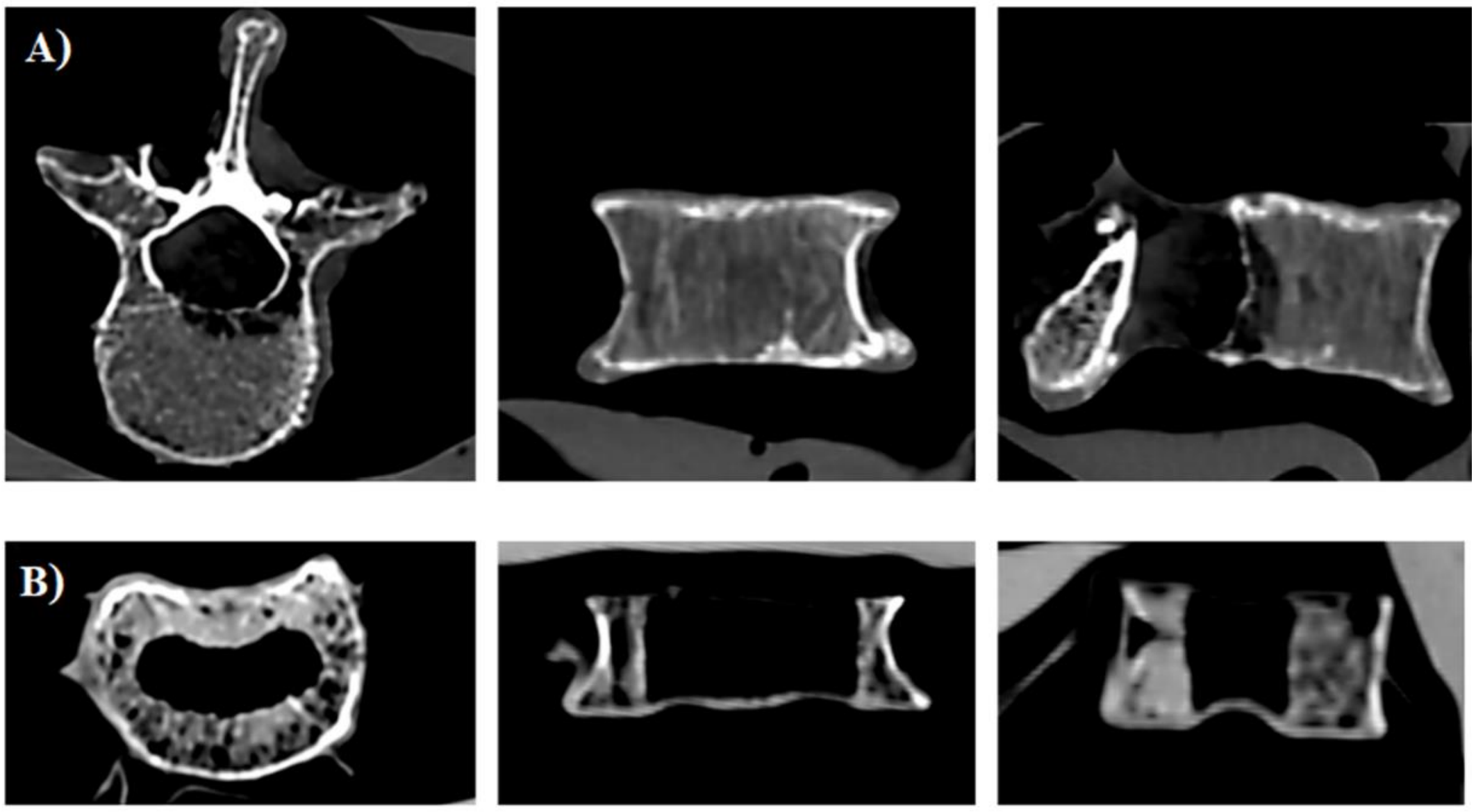

**Transversal plane** **Frontal plane** **Sagittal plane**

**Fig. 4.**

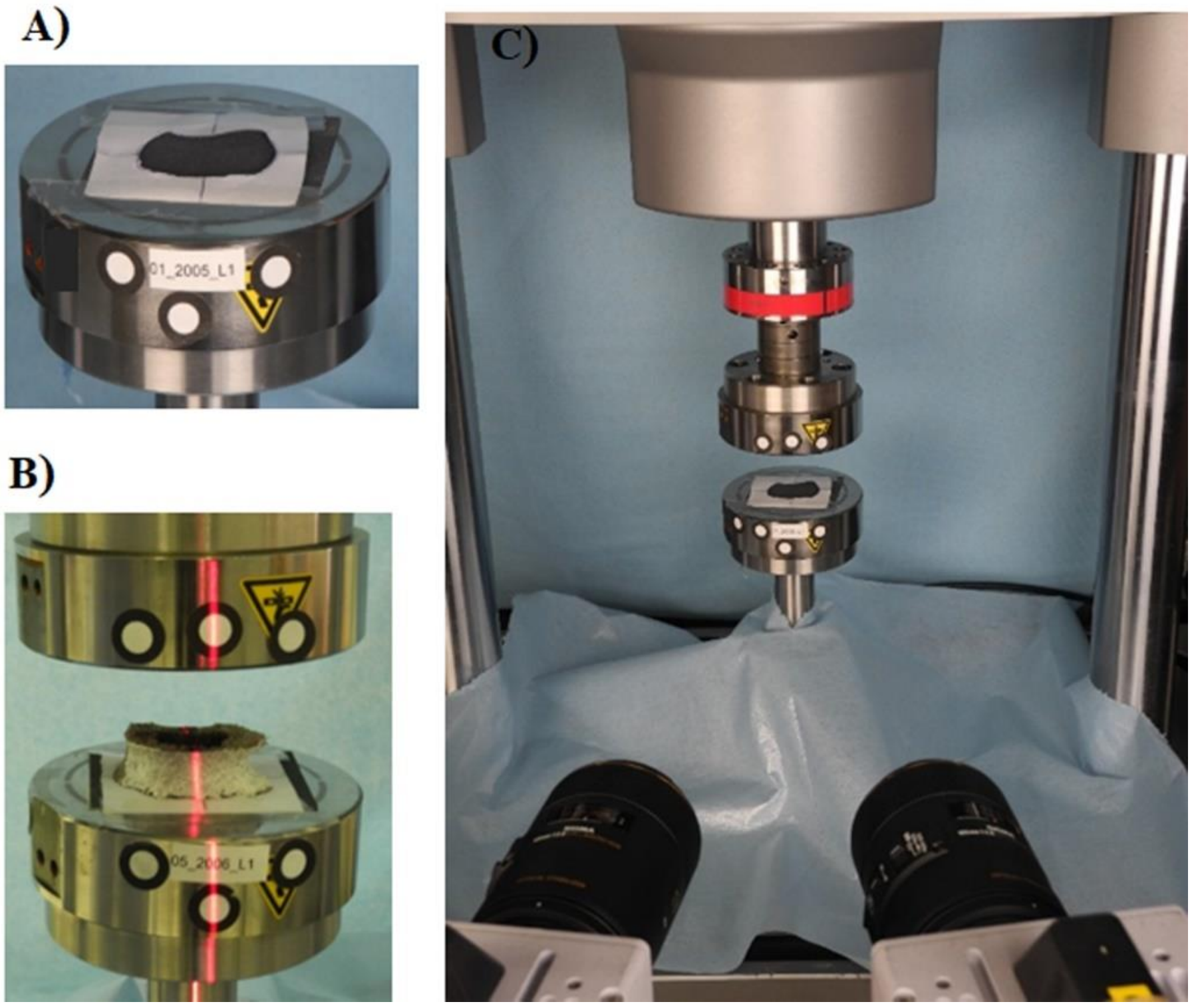

**Fig. 5.**

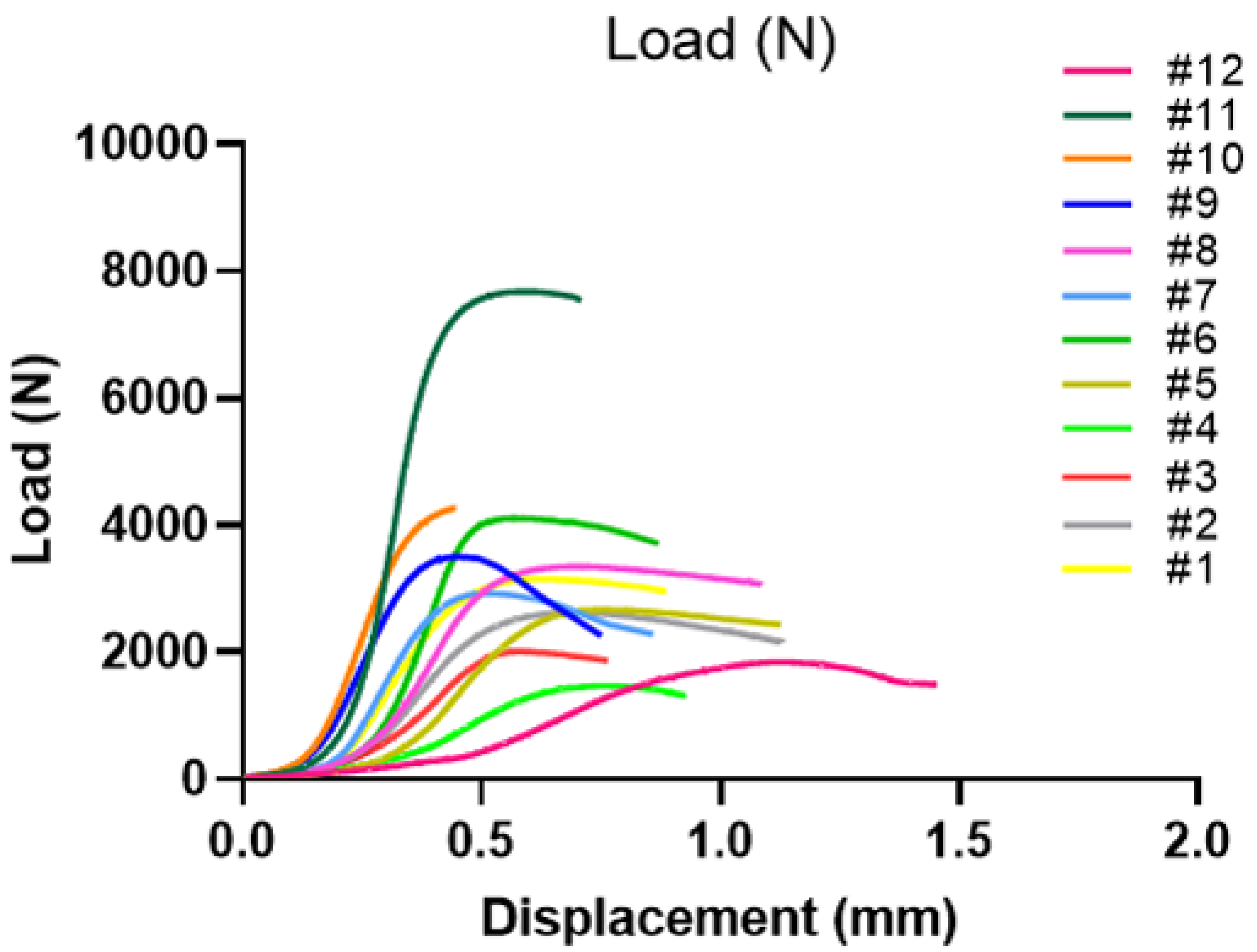

**Fig. 6.**

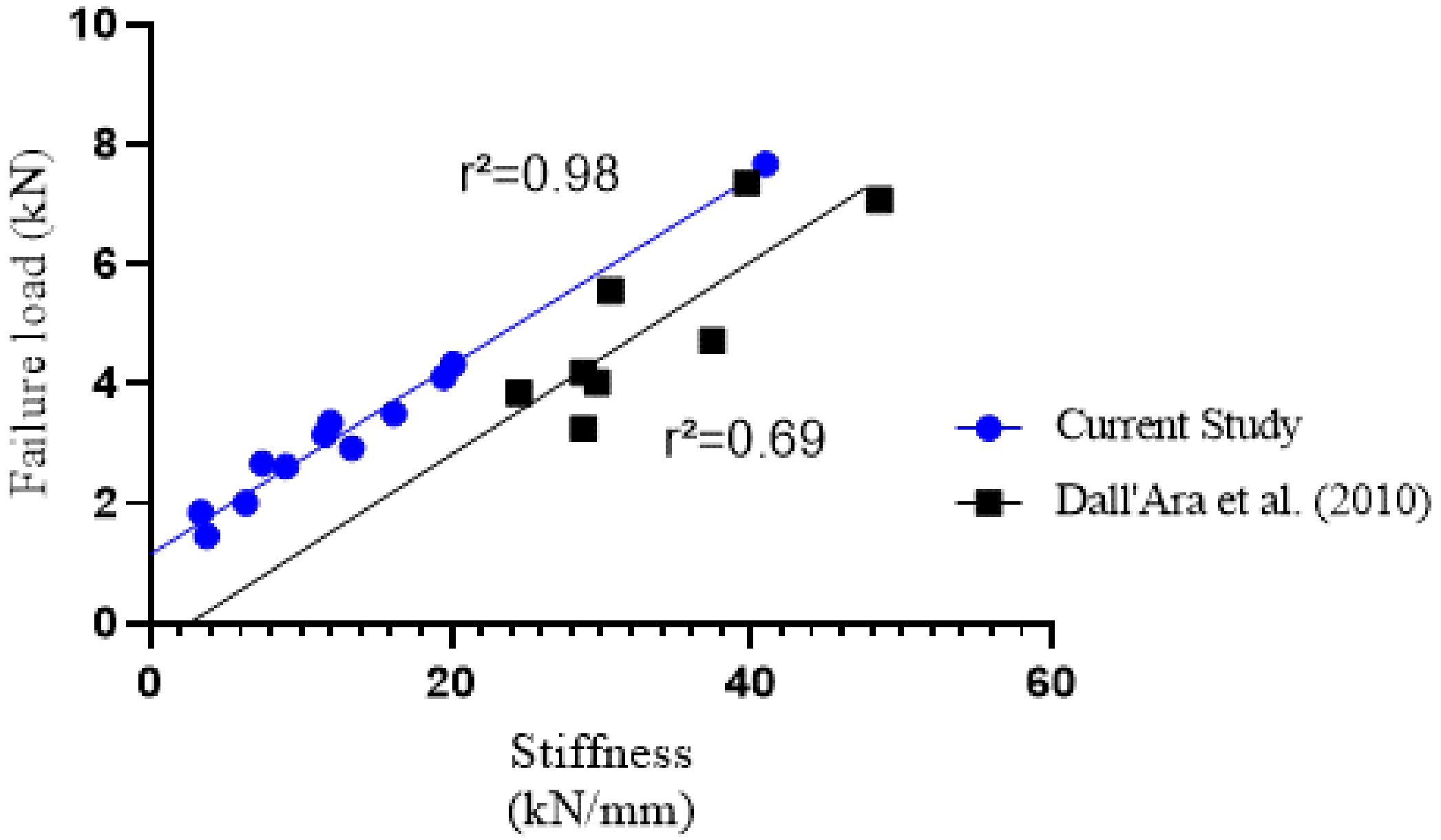

**Fig. 7.**

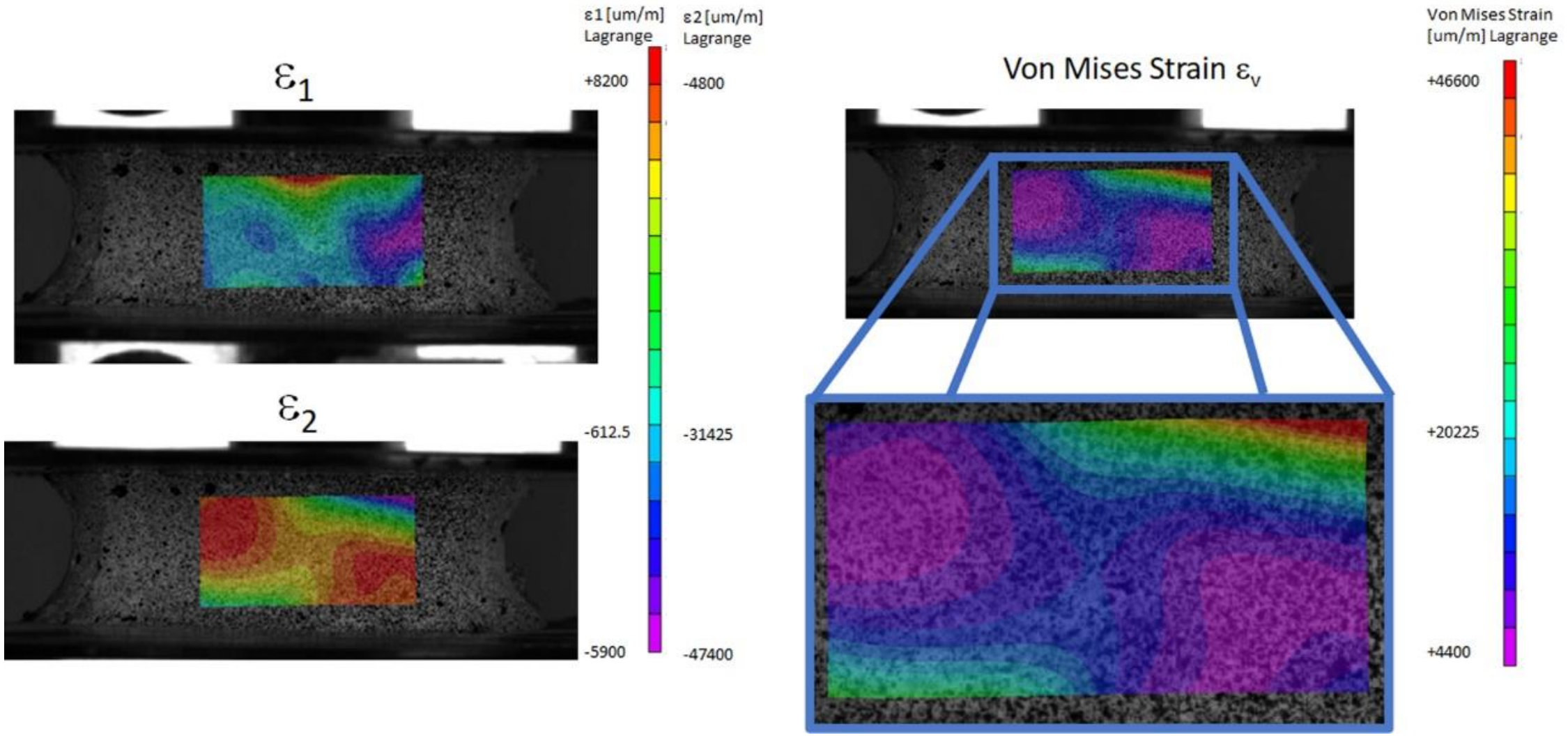

**Table 1.**

| ID | Sex | Age | pre-defect BMD from HRpQCT | pre-defect BMD from QCT | post-defect BMD from QCT | pre-defect BV/TV from QCT | Mean height after plate cutting | Defect size ratio measured in 2D (on mid-section) | Defect size ratio measured in 3D |
|---|---|---|---|---|---|---|---|---|---|
| | | | (g/cm$^3$) | (g/cm$^3$) | (g/cm$^3$) | (%) | (mm) | (%) | (%) |
| **1** | F | 72 | 0.163 | 0.118 | 0.046 | 32.4 | 13.7 | 32% | 27% |
| **2** | M | 95 | 0.067 | 0.028 | 0.01 | 11.3 | 14.8 | 25% | 26% |
| **3** | F | 93 | 0.066 | 0.014 | 0.050 | 3.7 | 13.2 | 26% | 25% |
| **4** | F | 81 | 0.075 | 0.032 | 0.017 | 11.7 | 15.3 | 27% | 28% |
| **5** | M | 88 | 0.119 | 0.067 | 0.048 | 18.0 | 17.0 | 20% | 28% |
| **6** | M | 73 | 0.158 | 0.098 | 0.053 | 27.3 | 12.4 | 24% | 31% |
| **7** | F | 93 | 0.110 | 0.058 | 0.015 | 17.7 | 13.6 | 29% | 31% |
| **8** | M | 84 | 0.124 | 0.062 | 0.030 | 18.9 | 16.8 | 33% | 29% |
| **9** | F | 85 | 0.114 | 0.064 | 0.016 | 19.5 | 13.8 | 25% | 26% |
| **10** | M | 80 | 0.059 | 0.011 | 0.021 | 3.1 | 14.8 | 23% | 27% |
| **11** | M | 73 | 0.158 | 0.092 | 0.046 | 23.3 | 13.2 | 27% | 29% |
| **12** | M | | 0.105 | 0.057 | 0.021 | 19.3 | 19.6 | 28% | 23% |
| **Mean** | | | **0.110** | **0.058** | **0.033** | **17.2** | **14.8** | **26%** | **27%** |
| **SD** | | | **0.037** | **0.033** | **0.016** | **8.69** | **2.06** | **0.04** | **0.02** |

**Table 2.**

| ID | Failure load (F) | Compression at failure load | Stiffness (k) | Mean von Mises strain at failure load | Mean max. principal strain at failure load (ε1) | Mean min. principal strain at failure load (ε2) | Apparent Max. Stress (F/Sapp) | Apparent Modulus (k*L/Sapp) |
|---|---|---|---|---|---|---|---|---|
| | (kN) | (mm) | (kN/mm) | (%) | (%) | (%) | (MPa) | (MPa) |
| 1 | 3.15 | 0.6 | 11.61 | 1.61 | 0.16 | -1.52 | 6.62 | 334 |
| 2 | 2.62 | 0.7 | 9.03 | 2.42 | 0.29 | -2.26 | 3.02 | 154 |
| 3 | 2.01 | 0.6 | 6.36 | 0.87 | 0.25 | -0.7 | 3.06 | 128 |
| 4 | 1.46 | 0.8 | 3.75 | 1.09 | 0.17 | -0.97 | 1.91 | 75 |
| 5 | 2.67 | 0.8 | 7.44 | 1.82 | 0.17 | -1.71 | 2.29 | 108 |
| 6 | 4.11 | 0.6 | 19.50 | 1.16 | 0.04 | -1.14 | 4.47 | 262 |
| 7 | 2.93 | 0.5 | 13.40 | 1.38 | 0.11 | -1.41 | 4.15 | 257 |
| 8 | 3.35 | 0.7 | 11.98 | 1.3 | 0.05 | -1.24 | 5.72 | 343 |
| 9 | 3.50 | 0.5 | 16.21 | 1.36 | 0.05 | -1.36 | 5.07 | 324 |
| 10 | 4.33 | 0.5 | 20.14 | 2.55 | 0.35 | -2.34 | 4.44 | 305 |
| 11 | 7.67 | 0.6 | 40.94 | 2.25 | 0.33 | -2.14 | 8.32 | 586 |
| 12 | 1.85 | 1.1 | 3.36 | 0.51 | 0.21 | -0.35 | 3.18 | 113 |
| **Mean** | **3.30** | **0.67** | **13.64** | **1.53** | **0.18** | **-1.43** | **4.35** | **249** |
| **SD** | **1.62** | **0.17** | **10.23** | **0.63** | **0.11** | **0.61** | **1.87** | **145** |

**Table 3.**

| | | Failure load (kN) | | | Stiffness (kN/mm) | | |
|---|---|---|---|---|---|---|---|
| **Study** | **Specimens** | **Mean** | **Median** | **Range** | **Mean** | **Median** | **Range** |
| **Dall'Ara et al. (2010)** | 8 L1 *(Intact)* | 5.00 | 4.46 | [3.25 – 7.35] | 33.52 | 30.19 | [24.54 – 48.54] |
| **Stadelmann et al. (2020)** | 4 L1 *(3 blastic, 1 lytic)* | 6.85 | 5.80 | [4.92 –12.27] | Not defined | | |
| **Stadelmann et al. (2020)** | 6 Lytic *(1T5, 1T6, 1T9, 1T11, 1L1, 1L3)* | 4.86 | 5.82 | [1.89 – 6.27] | Not defined | | |
| **Current study** | 12 L1 *(Created defects)* | 3.30 | 3.04 | [1.46 – 7.67] | 13.64 | 11.80 | [3.36 – 40.94] |